\documentclass[conference]{IEEEtran}
\IEEEoverridecommandlockouts
\usepackage{balance}
\renewcommand{\baselinestretch}{.99}
\usepackage[labelfont=bf]{caption}
\usepackage{subcaption} 
\usepackage[font=small,skip=3pt]{caption}
\usepackage[nodisplayskipstretch]{setspace}
\usepackage{booktabs}
\usepackage[table,xcdraw]{xcolor}
\usepackage{cite}
\usepackage{multirow}
\usepackage{amsmath,amssymb,amsfonts,mathtools}
\usepackage{hyperref}
\usepackage{cleveref}
\usepackage{algorithmic}
\usepackage{graphicx}
\usepackage{url}
\usepackage{bm}
\usepackage{cite}
\usepackage{mathtools,amssymb,amsfonts,amsmath}
\usepackage{bm}
\usepackage{algorithmic}
\usepackage{graphicx}
\usepackage{textcomp}
\usepackage{comment}
\usepackage{listings}
\usepackage{siunitx}
\usepackage{color}
\usepackage{enumitem}
\usepackage{booktabs}
\usepackage{gensymb}

\DeclareMathOperator*{\argmax}{arg\,max} %

\begin{document}
\setlength{\belowdisplayskip}{6pt} \setlength{\belowdisplayshortskip}{6pt}
\setlength{\abovedisplayskip}{6pt} \setlength{\abovedisplayshortskip}{6pt}
\setlength{\floatsep}{6pt}
\setlength{\textfloatsep}{6pt}
\setlength{\dbltextfloatsep}{6pt}
\setlength{\abovecaptionskip}{6pt}

\title{Sensitivity of a Chaotic Logic Gate}

\author{Noeloikeau F. Charlot, Daniel J. Gauthier ~\IEEEmembership{Member,~IEEE}}

\markboth{Journal of \LaTeX\ Class Files,~Vol.~14, No.~8, August~2021}%
{Shell \MakeLowercase{\textit{et al.}}: A Sample Article Using IEEEtran.cls for IEEE Journals}

\maketitle

\begin{abstract}
Chaotic logic gates or `chaogates' are a promising mixed-signal approach to designing universal computers. However, chaotic systems are exponentially sensitive to small perturbations, and the effects of noise can cause chaotic computers to fail. Here, we examine the sensitivity of a simulated chaogate to noise and other parameter variations (such as differences in supply voltage). We find that the regions in parameter space corresponding to chaotic dynamics coincide with the regions of maximum error in the computation. Further, this error grows exponentially within 4-10 iterations of the chaotic map. As such, we discuss the fundamental limitations of chaotic computing, and suggest potential improvements. Our Python simulation methods are open-source and available at \href{https://github.com/Noeloikeau/chaogate}{https://github.com/Noeloikeau/chaogate}.
\end{abstract}


\section{Introduction}
Chaotic systems have been shown to be able to implement any logical function \cite{chaos_infinite_functions}. Therefore, chaos can be used for computation \cite{chaos_computing}. In fact, a chaotic computer can perform faster and/or require fewer resources than a typical digital computer in certain cases \cite{molnr_2013_asymmetric,nonlinear_computation}. Moreover, the extreme sensitivity that characterizes chaotic systems may function as an inherent encryption mechanism, serving to obfuscate circuits and information \cite{chaos_ALU:rose}. As a result, chaotic circuits have the potential to offer faster, more secure, and more compact methods of computation than traditional hardware.

Here, we consider an established class of mixed-signal digital-analog computers built around a chaotic `core' (an analog nonlinear circuit) that interfaces with digital control logic \cite{hybrid_functions,integrated_chaogate}. This control logic is used to specify the initial condition of the chaotic core and the amount of time that it is allowed to operate or `iterate.' After the specified amount of time or number of iterations, the analog state of the core is binarized and converted into a digital value by a comparator circuit (register, Schmitt trigger, \textit{etc.}), the result of which is used in the output of the computation.

In practice, different measurement times are used to select different logical functions. This is because the state of the chaotic core changes over time, with a tendency to eventually visit the full range of states that the system can occupy - a property known as ergodicity \cite{ergodic}. Hence, as time changes, a different logical function tends to describe the mapping from the initial condition to the current Boolean state. As a result, a chaotic computer is effectively described by a look-up-table (LUT) listing which function is implemented given an iteration number, initial condition, and control signal.

However, if repeated measurements of the chaotic system vary for a fixed input (such as due to thermal noise, fluctuating power supply, \textit{etc.}) then the computation is non-deterministic, as the same input may yield different outputs. This changes the LUT entries describing the computer from functions to distributions of functions, each occurring with some probability dependent upon the characteristics of the device and its environment. As such, accounting for the sensitivity of the Boolean output to the operational parameters of the analog circuit (such as its control voltage) is essential in realizing deterministic chaotic computation in practice \cite{fault_tolerant,chaos_ALU:rose}. However, to the best of our knowledge, such a sensitivity analysis has not been thoroughly presented in the literature.

Here, we perform a fundamental sensitivity analysis on a single chaotic logic gate or chaogate \cite{ditto2010chaogates}. The core of the chaogate is composed of three coupled transistors forming an analog nonlinear circuit \cite{chaos_discrete_time_generator:dudek,chaogate_circuit}. Recently, we introduced a comprehensive simulation and optimization framework for this system \cite{chaogate_optimization}. Using this framework, we analyze the properties of the chaogate, namely its Lyapunov exponent, parametric sensitivity, and noise sensitivity (error rate). Our unique contributions follow:

\begin{itemize}
    \item We show that chaogate noise and parameter sensitivities coincide with a positive Lyapunov exponent, and that error accumulates exponentially within $10^{1}$ iterations.
    \item Our results demonstrate that chaogates are exponentially sensitive to small perturbations of any kind, suggesting that there exists a trade-off between chaos and deterministic computation in experimental systems. 
\end{itemize}

The paper is organized as follows. In Sec. \ref{methods} we describe the chaogate simulation and precisely define `chaos' and `sensitivity'. In Sec. \ref{results} we analyze these properties across an optimal region of parameter space and discuss the results. In Sec. \ref{conclusion} we conclude with future research. 

\section{Design and Methodology}
\label{methods}
\label{design}

\begin{figure}[htbp]
\centerline{\includegraphics[width=3.5in]{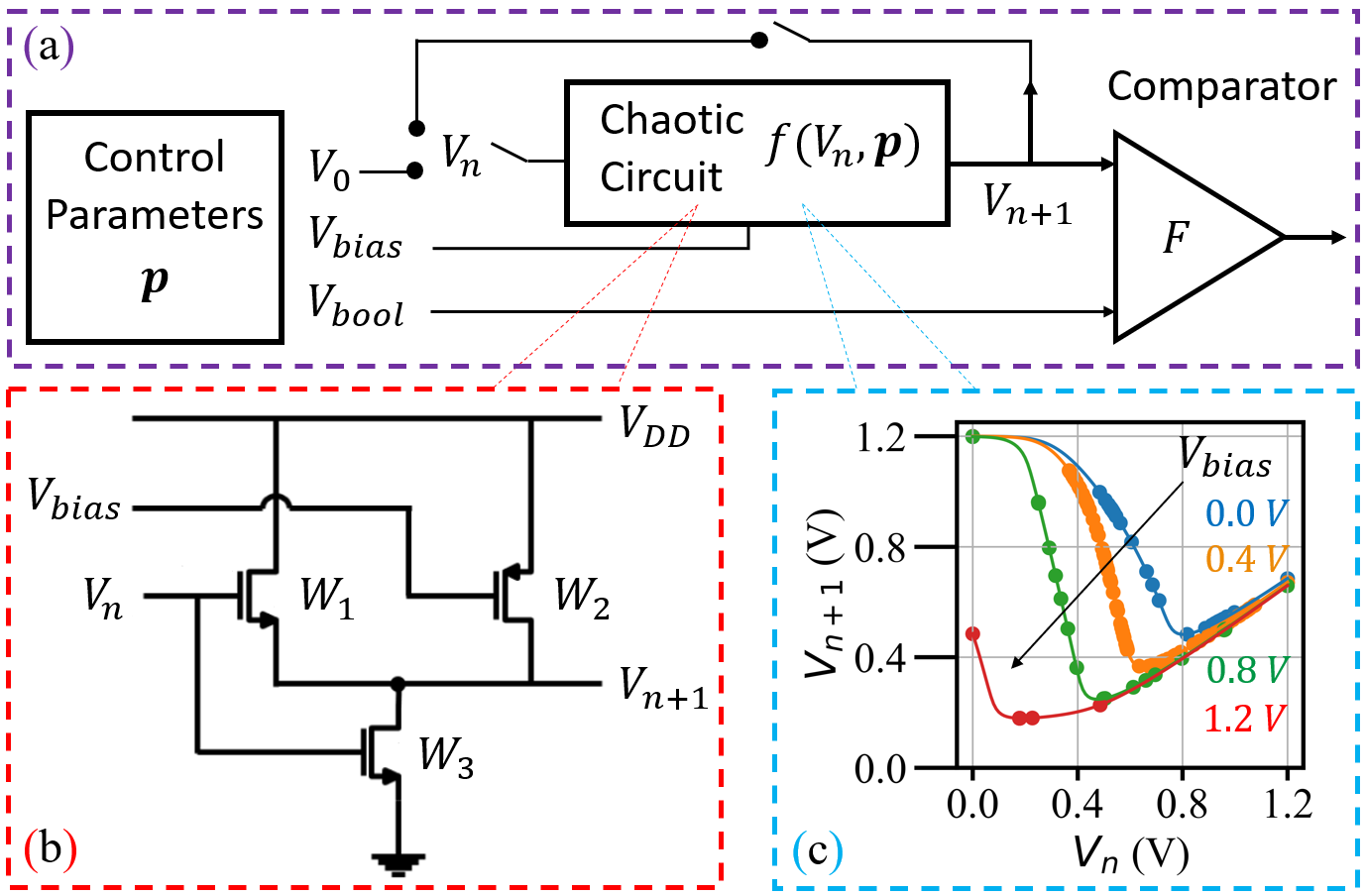}}
\caption{Chaogate circuit diagram. (a): Sample-and-hold design that iterates a map $f$ taking the input voltage $V_{n}$ to the output voltage $V_{n+1}$, which is used as the next input. (b): Chaotic core containing 3 MOSFETs of widths $W_{1}$ and $W_{3}$ (p-type) and $W_{2}$ (n-type). (c): Voltage sweep showing the DC transfer functions (solid curves $f$) at various $V_{bias}$ that are used to iterate the map (scattered points $V_{n}$).}
\label{chaogate}
\end{figure}

Each chaogate (Fig. \ref{chaogate} (a)) implements a function $f$ that maps an input voltage $V_{n}$ to an output voltage $V_{n+1}$, given a fixed set of parameters $\bm{p}$. Control logic initially holds a chaotic circuit (Fig. \ref{chaogate} (b)) to the input voltage $V_{0}$. The circuit equilibrates and produces an output voltage $f(V_{0})=V_{1}$. This is sampled by the control logic and used as the next input (producing an output $f(V_{1})=V_{2}$, and so on). The `sample and hold' design repeats this process $N$ times in a procedure known as iteration. At each iteration $n$, the output voltage is passed through a comparator $F$, which produces a bit used in the result of the computation. 

\begin{table}[ht]
\centering
\caption{Chaogate netlist parameters $\bm{p}$ and their values used in simulation, which are either fixed, or listed in [start,stop,step] format.}
\begin{tabular}{|c c|}
\hline
 Parameter & Value(s)\\
 \hline\hline
  Control Voltage & $V_{bias}\in[0,1.2,0.01]$ V\\
 \hline
  Power Supply & $V_{DD}\in[1.15,1.25,0.001]$ V\\
 \hline
  MOSFET Widths & $\bm{W}\in[65,2000,65]^{3}$ nm\\
 \hline
  MOSFET Lengths & $\bm{L}=(65,65,65)$ nm\\
 \hline
  Temperature & $T=25$ $^\circ$C\\
 \hline
  Initial Condition & $V_{0}=0$ V\\
 \hline
  Logic Threshold & $V_{bool}=0.6$ V\\
 \hline
\end{tabular}
\label{parameters}
\end{table}

The chaogate circuit simulation is specified by a list of parameters $\bm{p}$, and the resulting dynamics give rise to the form of $f$. Mathematically, the chaogate acts as a one-dimensional map at each point in parameter space $f|_{\bm{p}}:V_{n}\mapsto V_{n+1}$, and its functional form is found by iterating a recurrence relation
\begin{equation}
    V_{n+1}(\bm{p}) = f(V_{n},\bm{p}).
    \label{map}
\end{equation}

Parameters correspond to the physical operating conditions and circuit characteristics in which iteration is performed, and are assumed to be held fixed during simulation. This assumption is reasonable due to the sample-and-hold nature of the digital control logic (Fig. \ref{chaogate} (a)), which causes the circuit to function at its DC operating point. This suppresses transient fluctuations in all but the final output voltage supplied to the next stage. As such, the logical output of the chaogate is a function of only this voltage and the Booleanization threshold
\begin{equation} \label{bool}
    F(V_{n},\bm{p})=\begin{cases} 
      0 & f(V_{n},\bm{p}) < V_{bool}, \\
      1 & f(V_{n},\bm{p}) \geq V_{bool},
   \end{cases}
\end{equation}
where $V_{bool}=0.6$ V is the intermediary logic threshold, set to halfway between the range of $V_{n}\in[0,1.2 \text{ V}]$. We simulate the circuit using standard `NgSpice' tools, and perform a voltage sweep over $V_{n}$ in order to determine the functional form of $f$, which we then interpolate for iteration (Fig. \ref{chaogate} (c)). See \cite{chaogate_optimization} and the `chaogate' Python package documentation for details.

The behavior of $F$ in both time ($n$) and parameter space ($\bm{p}$) forms the basis of any subsequent computation. Whether or not $F$ is chaotic depends heavily on the particular value of the parameters, and in practice control logic is introduced which modifies these parameters and implements the sample-and-hold/iteration process. Additionally, multiple chaogates are typically combined to form higher-order logical functions. However, the implementation of this control and combinational logic is design-dependent. For generality we study the behavior of $F$ by varying $\bm{p}$ directly, without extraneous logic. The output of $F$ at different $\bm{p}$ can then be combined as needed to form higher-order logical functions, which we do not consider here.

\subsection{Lyapunov Exponent}
\label{lyapunov}
Chaos in a one-dimensional map is defined by a positive Lyapunov exponent $\lambda>0$ \cite{schuster2006deterministic}. $\lambda$ is an exponential measure of the rate at which a sequence diverges in time. The \textit{local} Lyapunov exponent $\lambda_{n}$ at iteration $n$ of a chaogate with fixed parameters $\bm{p}$ and initial condition $V_{0}\in\bm{p}$ is the logarithm of the absolute value of the derivative
\begin{equation}
    \lambda_{n}(\bm{p})=\ln\bigg|\frac{\partial f}{\partial V_{n}}(V_{n},\bm{p})\bigg|.
\end{equation}
The (asymptotic) Lyapunov exponent $\lambda$ is then defined as
\begin{equation}
    \lambda(\bm{p}) = \lim_{N\rightarrow\infty}\frac{1}{N}\sum_{n=1}^{N}\lambda_{n}(\bm{p}).
    \label{lambda}
\end{equation}
We find the limit is well approximated by $N=1000$ and fix this quantity throughout. Further, we find that $\lambda$ is asymptotically independent of $V_{0}$, indicating that the chaogate is an ergodic system \cite{fault_tolerant}. As such, we fix $V_{0}=0$ V and discard the first $100$ transient iterations in our calculations of $\lambda$, which we find yields representative results. 

\subsection{Parameter Sensitivity}
\label{parametersensitivity}
The capacity for the chaogate to act as a deterministic computer is directly related to whether its logical output changes in the presence of small fluctuations on the control parameters. In reality, parameters fluctuate under experimental operating conditions, and this can lead to two bitstreams produced by the same chaogate with the same input diverging over time. Thus, the \textit{parametric sensitivity} $\sigma_{n}$ of the $n$'th iteration can be characterized by the average distance between Boolean states at nearby points in parameter space
\begin{equation}
    \sigma_{n}(\bm{p},\hat{\mathcal{P}}) = \sqrt{\frac{1}{||\hat{\mathcal{P}}||}\sum_{i\in \hat{\mathcal{P}}}\bigg(\frac{\partial F}{\partial p_{i}}(V_{n},\bm{p})\Delta p_{i}\bigg)^{2}},
    \label{beta}
\end{equation}
where $\hat{\mathcal{P}}$ is the set of parameters over which the sensitivity is calculated, $||\hat{\mathcal{P}}||$ is the number of elements in this set, and $\Delta p_{i}$ is the increment value of parameter $i$ (see Table \ref{parameters}). In analogy to the Lyapunov exponent, we define the asymptotic parametric sensitivity $\sigma=(1/N)\sum\sigma_{n}$. 

\subsection{Noise Sensitivity}
\label{noisesensitivity}
In addition to parametric sensitivity, the chaogate also possesses noise sensitivity. Here we consider noise to be an additive voltage fluctuation on the output terminal, modifying Eq.~\ref{map} to read
\begin{equation}
    V_{n+1}(\bm{p}) = f(V_{n},\bm{p})+\mathcal{N}(0,\nu^{2}),
\end{equation}
where $\nu\in\bm{p}$ is the standard deviation of the output voltage fluctuation in volts, drawn from the Normal distribution $\mathcal{N}$ with zero mean at each iteration. The \textit{noise sensitivity} $\varepsilon_{n}$ of the $n$'th iteration is obtained by averaging the pairwise differences between $N_{r}=1000$ repeated measurements
\begin{equation}
    \varepsilon_{n}(\bm{p}) = \frac{2}{N_{r}(N_{r}+1)}\sum_{r=1}^{N_{r}}\sum_{r'=r+1}^{N_{r}}\bigg|F^{n_{r}}(\bm{p})-F^{n_{r'}}(\bm{p})\bigg|,
    \label{sigma}
\end{equation}
where $F^{n_{r}}$ is the $n$'th iterate of the $r$'th repetition, \textit{i.e.}, the $n$'th bit of Boolean string $r$ obtained by iterating the chaogate map with parameters $\bm{p}$, where each string is specified by the unique noise vector drawn over all timesteps. Hence, $\varepsilon_{n}$ is a local measure of the probability that an error occurs due to noise. In our calculations of $\lambda$ and $\sigma$, we take $\varepsilon(\nu=0)=0$ so that the results are purely deterministic. Similarly, we define the asymptotic noise sensitivity as $\varepsilon=(1/N)\sum\varepsilon_{n}$.

Here, we use the value $\nu=0.001$ V, obtained via the `.noise' function in NgSpice, which calculates the total output noise from all elements of the chaogate circuit on the output voltage terminal $V_{n+1}$ (see Fig. \ref{chaogate}). To estimate the total noise, we superimpose an arbitrarily small AC sinusoid on the supply voltage $V_{DD}$. This noise source affects the entire circuit, and we integrate over the frequency range $[1,10^{10}]$ Hz. This provides a conservatively large estimate for the output noise in the sample-and-hold circuit design.

\subsection{Optimization and Chaotic Sensitivities}
\label{optimization}
In order to maximize the design space for a chaotic computer, we search for the largest region of parameter space over which the dynamics are chaotic. In practice, the three MOSFET widths $\bm{W}$ are the only parameters required for tape-out after the technology node $\bm{L}$ has been chosen. Hence, we select an optimal width $\bm{W}_{opt}$ defined as maximizing the area of the chaotic region $\mathcal{C}^{\star}$ in the $V_{bias}-V_{DD}$ plane (two typical control parameters in an experiment)
\begin{equation}
    \mathcal{C}^{\star}(\bm{W})=\{(V_{bias},V_{DD},\bm{W})\in\mathcal{G}\mid\lambda(V_{bias},V_{DD},\bm{W})>0\},
\end{equation}
\begin{equation}
    \bm{W}_{opt}=\argmax_{\bm{W}\in\mathcal{G}}\bigg( ||\mathcal{C}^{\star}(\bm{W})|| \bigg),
\end{equation}
where $\mathcal{G}$ is the Cartesian product of the parameter values defined by Table \ref{parameters} (\textit{i.e.}, a grid search). In practice we find that approximately half the width space is non-chaotic, emphasizing the importance of parameter optimization on building a chaotic computer. As a result of this search, we find $\bm{W}_{opt}=(725,65,1365)$ nm and fix this quantity throughout. However, we find that other chaotic regions of parameter space (obtained through random selection and linear combinations of the sensitivity measures $a\lambda-b\sigma-c\varepsilon$) display similar chaotic sensitivities, defined as follows.

We develop a characteristic measure of the chaogate sensitivity at each time by averaging over the chaotic region of parameter space $\mathcal{C}^{\star}$. This gives an expected value for the local sensitivity measures $\chi_{n}\in\{\lambda_{n},\sigma_{n},\varepsilon_{n}\}$ of a chaotic timeseries. In practice we find these data are noisy, and so employ a moving average to improve the quality of the signals, which we then fit analytically to a saturating exponential. We define the \textit{chaotic sensitivities} $\bar{\chi}_{n}$ at each iteration as
\begin{equation}
    \bar{\chi}_{n}(\mathcal{C}^{\star}(\bm{W}_{opt}))=\frac{1}{(N_{T}+1)||{\mathcal{C}^{\star}}||}\sum_{\bm{p}\in\mathcal{C}^{\star}}\sum_{n'=n-N_{T}/2}^{n+N_{T}/2}\chi_{n'}(\bm{p}),
    \label{average_sensitivity}
\end{equation}
where $N_{T}=2$ is the size of the window (nearest-neighbor).

\section{Results and Discussion}
\label{results}
\begin{figure}[htbp]
\centerline{\includegraphics[width=2.75in]{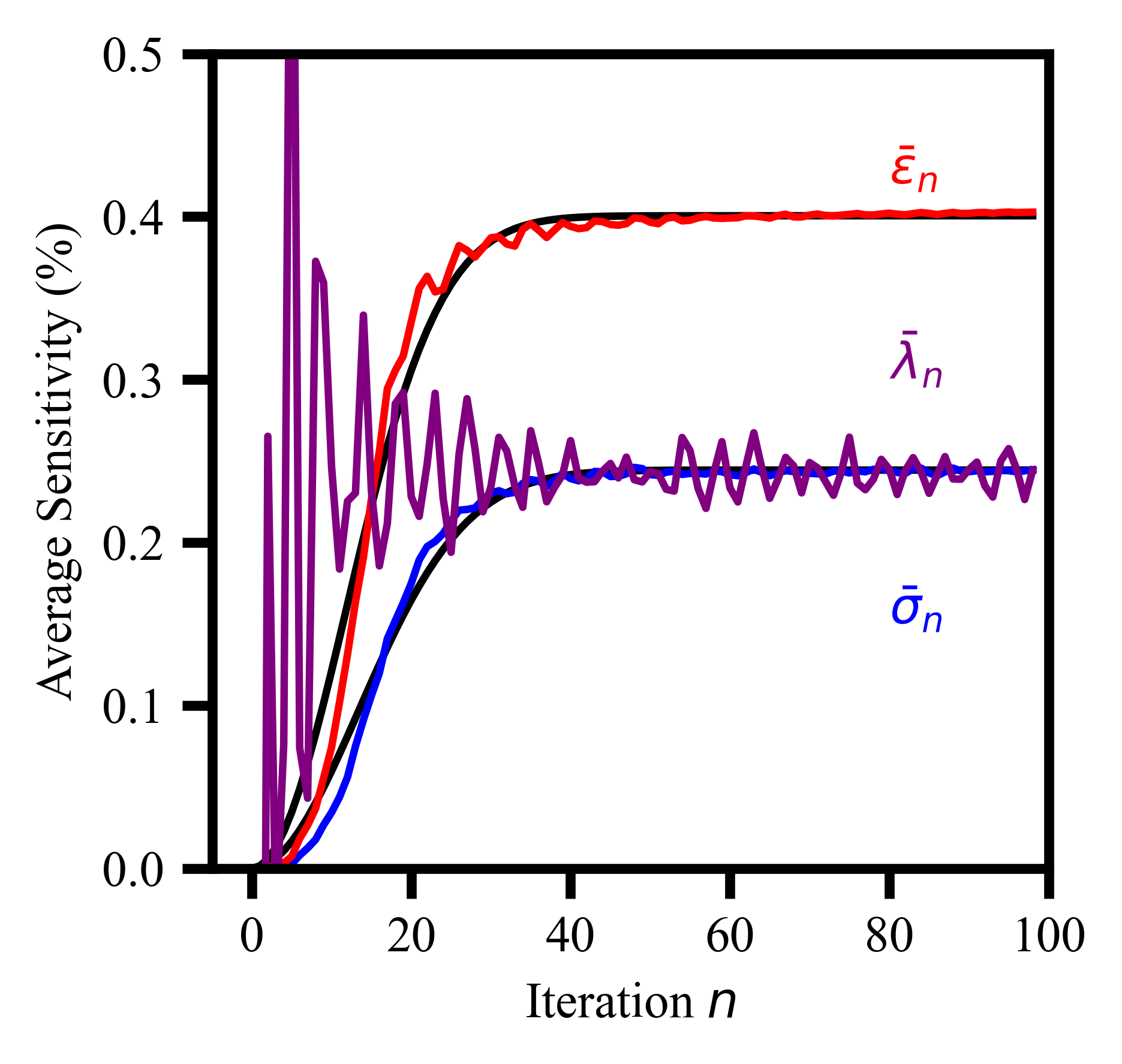}}
\caption{Chaotic sensitivities vs. number of iterations. The black curves underneath $\bar{\varepsilon}_{n}$ and $\bar{\sigma}_{n}$ are fits to a saturating exponential function of the form $y(n)=A(1-e^{-Cn^{2}})$. The fit coefficients are $A_{\varepsilon},C_{\varepsilon}=(0.401,0.004)$ and $A_{\sigma},C_{\sigma}=(0.244,0.003)$. The early fluctuations of $\bar{\lambda}_{n}$ have been truncated for visibility.}
\label{average}
\end{figure}

Shown in Figure \ref{average} are the chaotic sensitivities of the chaogate over the first 100 iterations. As can be seen, $\bar{\varepsilon}_{n}$ and $\bar{\sigma}_{n}$ are well fit to a saturating exponential function. These results show that chaogates are exponentially sensitive to perturbations of any kind - whether they are dynamical ($\lambda$), parametric ($\sigma$), or stochastic ($\varepsilon$). 
\begin{figure*}[t]
    \centering
    \includegraphics[width=7.0in]{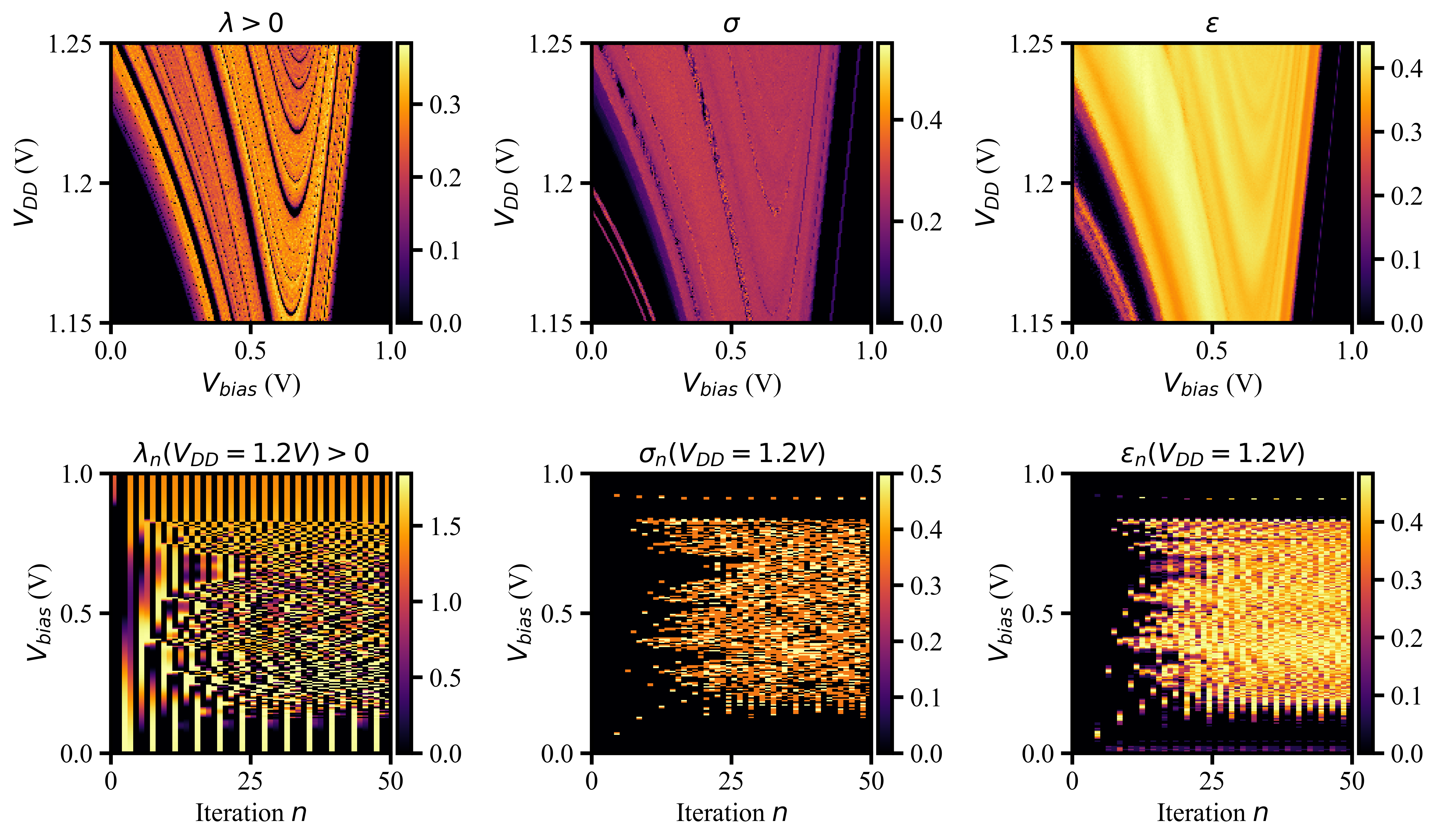}
    \caption{Heat maps of the asymptotic (top) and local (bottom) chaotic Lyapunov exponent $\lambda>0$, and parameteric $\sigma$ and noise $\varepsilon$ sensitivities. Here $\sigma$ is calculated over $\hat{\mathcal{P}}=\{V_{bias},V_{DD}\}$. Similar results ($\varepsilon \sim \lambda \sim \sigma$) are observed for other parameter slices.}
    \label{sensitivity}
\end{figure*}

From Fig. \ref{average}, we note the following two observations: 1. $\bar{\sigma}_{n}\approx\bar{\lambda}_{n}$ (the parametric sensitivity approximates the Lyapunov exponent), and 2. $\bar{\varepsilon}_{n}>\bar{\lambda}_{n}$ (the noise sensitivity is greater than the Lyapunov exponent). The first observation indicates that the rate at which the chaogate sequence divergences in time is equivalent to the way in which it diverges in parameter space. This is perhaps surprising as the quantities are calculated from derivatives over independent axes. However, the ergodicity of the chaogate implies that `time'-averages ($\lambda$) and `space'-averages ($\sigma$) are approximately equal under certain conditions \cite{ergodic}.

The second observation indicates that the effects of noise eventually overwhelm the system, becoming the dominant source of error ($A_{\varepsilon}>A_{\sigma}$ in Fig. \ref{average}). This is also demonstrated by the slightly shorter saturation time for the noise sensitivity (defined as the time required for the function to reach 1/e of its maximum value $n_{\chi}=\sqrt{-\ln(1-e^{-1})/C_{\chi}}$), here given by $n_{\varepsilon}=11.29<n_{\sigma}=12.57$. Thus, the transient saturation timescale is on the order of $10^{1}$ iterations for both sensitivities. As a lower-bound, we repeat this analysis using nanovolt noise amplitude $\nu\approx10^{-9}$ V, and observe similar functional behavior of $\bar{\varepsilon}_{n}$, but shifted forward in time by $40$ iterations. Thus, exponential error accumulation ensures the effects of noise can be forestalled but not eliminated.

Figure \ref{sensitivity} depicts heat maps of the chaogate measures over the supply-bias voltage plane (top), and the iteration-bias voltage plane (bottom). As can be seen, the chaotic regions $\lambda>0$ correspond with the regions of maximum noise \textit{and} parametric sensitivity. Additionally, certain non-chaotic regions still possess nonzero sensitivity, especially when surrounded by chaos. This demonstrates that chaogates existing on the `edge of chaos' accrue a nonzero error rate that approaches its maximum inside the fully chaotic regions. Hence, these error rates are inseparable from the chaotic dynamics, and there exists a design trade-off between the two.

\begin{figure}[htbp]
\centerline{\includegraphics[width=2.75in]{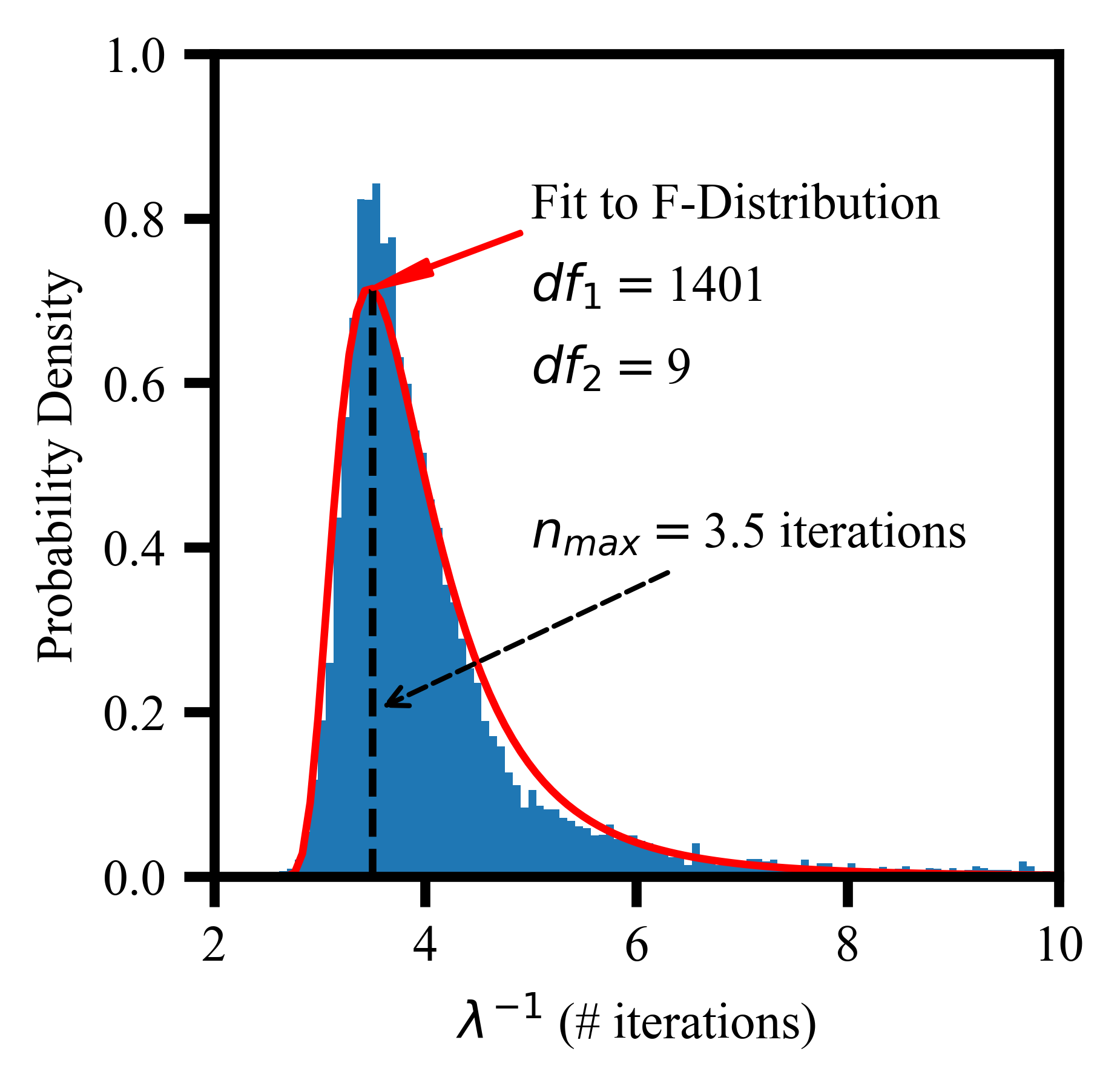}}
\caption{Lyapunov times $1/\lambda$ for the chaotic regions of parameter space $\lambda>0$. The most probable error rate is $e^{-1}/n_{max}$. Probability densities are constructed in Python using `numpy.histogram(...,density=True)' and fit using `scipy.stats.f.fit'.}
\label{times}
\end{figure}
Finally, we characterize the timescale over which the chaogate diverges by a factor of $1/e$. This is known as the Lyapunov time, and it is given by the reciprocal of the Lyapunov exponent $1/\lambda$. Shown in Fig. \ref{times} are the distribution of Lyapunov times over the chaotic parameter space. These are the asymptotic error rates of the system. The data are sharply peaked at $n_{max}=3.5$, meaning that the most frequently observed timescale over which an error grows by a factor of $e$ is 3.5 iterations. The data are bounded by the transient timescale ($10^{1}$ iterations; saturation time in Fig. \ref{average}). This means that errors typically grow exponentially within 4-10 iterations of the chaotic map.

Our findings are corroborated by experimental measurements of the error rate of a chaogate as a function of number of iterations, which show that some logical functions become random within the first 10 iterations (see Table 1 in Ref. \cite{integrated_chaogate}). These measurements match the predicted timescales, suggesting that our simulations are sufficiently detailed to predict the ensemble statistics of physical chaogates. 

\section{Conclusions and Future Research}
\label{conclusion}
In summary, we show that 3-transistor chaotic logic gates are exponentially sensitive to perturbations of any kind, whether these are noise-induced, or due to fluctuating control parameters. We find that these sensitivities arise from a positive Lyapunov exponent, and that sensitivity is therefore a fundamental property of chaotic systems. We go on to estimate chaogate error rates at $e^{-1}$ every 4-10 iterations. Therefore, we conclude that chaogate computing applications are limited, and propose that alternative designs should be explored - including improved error correction mechanisms, and fully analog, non-deterministic computation. We also suggest that the observed exponentially scaling parametric sensitivity implies a specific application for chaogates in cryptography. 

Extreme sensitivity to manufacturing variations characterizes Physically Unclonable Functions (PUFs) \cite{gao_2020_physical}. PUFs act as `digital fingerprints' by transforming the physical differences between devices (such as differences in transistor width) into bits used for identification. Hence, chaogates may be applicable as PUFs due to their exponential parametric sensitivity. We note that $\varepsilon$ is mathematically identical to the typical measure of reliability $\mu_{intra}$ in PUF literature, and a similar extension exists for $\sigma$ and the uniqueness measure $\mu_{inter}$. 

More robust error correction mechanisms could improve chaogate reliability (reduce noise error) by selecting for outliers in parameter space, and attempting to preserve long-lived chaotic transients $(\lambda^{-1}>10\text{ iterations})$ robust to noise. These outliers possess small but nonzero positive Lyapunov exponents, yielding large Lyapunov times with potential applications in computing. However, noise can dramatically reduce the lifetime of these transients \cite{lohmann_2017_transient}, and balancing these effects is a challenging optimization problem for future research.

Alternatively, using an autonomous circuit as the chaotic core could potentially leverage these sensitivities as properties of the dynamics. Rather than a sample-and-hold circuit, a transmission line could be used to directly connect the input terminal to the output terminal. This produces an autonomous system in which the output of the computation is mapped directly onto the analog attractor of the AC circuit dynamics. The effects of noise could potentially hasten the approach of the chaotic transient toward the attractor, converging on a solution within a probabalistically bounded timeframe. The analysis of such a circuit is the subject of future research. 

\section*{Acknowledgments and Disclaimer}
This material is based on research sponsored by Air Force Research Lab (AFRL) under agreement number FA8650-19-1-1741. The U.S. Government is authorized to reproduce and distribute reprints for Governmental purposes notwithstanding any copyright notation thereon. The views and conclusions contained herein are those of the authors and should not be interpreted as necessarily representing the official policies or endorsements, either expressed or implied, of Air Force Research Lab (AFRL) or the U.S. Government. The authors would also like to acknowledge helpful discussions with Domenic Forte and Rabin Acharya.
\vfill\break

\bibliographystyle{ieeetr}
\balance
\bibliography{references}
\end{document}